\documentclass[preprint]{elsart}
\usepackage{natbib}
\usepackage{graphicx}
\usepackage{amssymb}

\journal{}%Journal of Theoretical Biology}

\begin{document}

\begin{frontmatter}
\title{A mathematical model for period-memorizing behavior 
in {\it Physarum} plasmodium}
\author{Masashi Tachikawa}
\address{ERATO Complex Systems Biology Project, 
           JST, 3-8-1, Komaba, Meguro-ku, Tokyo 153-8902, Japan.}
\date{, 2009}
\ead {mtach@complex.c.u-tokyo.ac.jp}


\begin{abstract}
A mathematical model to describe period-memorizing behavior 
in {\it Physarum} plasmodium are reported.
In constructing the model, we first examine the basic characteristics
 required for the class of models, then create a minimal linear model to 
fulfill these requirements. 
We also propose two modifications of the minimal model,
nonlinearization and noise addition, which improve the reproducibility
 of experimental evidences.
Differences in the mechanisms and in the reproducibility of
experiments between our models and the previous models are discussed.
\end{abstract}
\maketitle
\end{frontmatter}

\section{Introduction}

The plasmodium of {\it Physarum polycephalum} is a huge 
unicellular organism with amoeboid movement.
It is differentiated into an advancing front zone 
and a rear region that is composed of a network of 
protoplasmic veins.
The concentrations of many chemical components in its 
cytoplasm, such as Ca$^{2+}$\citep{rid76,yos81a}, ATP\citep{yos81b}, 
H$^+$\citep{nak82}, cAMP\citep{ued86}, and NADH\citep{mor87}, 
display mutually entrained oscillations in a period of 1-2 minutes.
These oscillations also involve the contraction-relaxation cycle of 
cytoplasmic actomyosin\citep{woh79}, which generates rhythmic shrinkage of 
protoplasmic veins and the active transport of cytoplasm via shuttle 
streaming in the vein network.
The streaming of cytoplasm pushes the front zone to spread out, 
and the solation of cytoplasmic gel in the zone occurs 
at the same time; thus the plasmodium migrates.
This shuttle transport of cytoplasm also coordinates 
the phase relations among local oscillatory dynamics 
and lets a huge plasmodium behave as a unified individual 
organism\citep{yos78}.

Because of its particular behaviors, 
the plasmodium of {\it P. polycephalum} has been intensively 
examined as a model organism for the study of the cooperative phenomena 
in biological systems.
Many theoretical approaches have also been reported, in which 
the plasmodium is modeled as spatially extended oscillatory medium
\citep{tep91,hya99}
or coupled oscillator system\citep{aka99,ter05,ode84,smi94}.
A combination of the results of experimental 
and theoretical studies have revealed 
a lot of interesting features about plasmodium's behaviors,
such as 
entrainment to environmental stimulation\citep{miy92}, 
mutual entrainment\citep{yos78},  
symmetry-induced patterns\citep{tak01,tak06}, 
maze-solving\citep{nak00}, and computing\citep{aon08}.

Recently, another interesting behavior --
memorization of the period of periodic stimulation --
has been reported by \cite{sai08}.
In that study, the {\it Physarum} plasmodia
were exposed to dry and cold condition 
(dry stimulation) 
three times periodically
with periods $T=30$-$90$ minutes.
The stimulation reduced the locomotion speeds of 
the {\it Physarum} plasmodia.
After the last stimulations, 
spontaneous slowdowns of locomotion speeds were observed.
The time intervals between the last stimulations and spontaneous 
slowdowns coincided with the stimulation periods.
This phenomenon was referred to as a spontaneous in-phase slowdown
(SPS). 
The number of SPS occurrences variated from none to
three, depending on the organism.
Additionally at $T=60$min after SPS stopped, 
one-time stimulation induced the spontaneous slowdowns 
after an equal period of time from the additional stimulation,
which is referred to as SPS after one disappearance (SPSD).  
SPSD was not observed at $T=40$ min or $T=80$ min.
These experiments clearly indicate that the {\it Physarum} plasmodium  
memorizes the period of the periodic stimulation. 
Furthermore, since the dry cold conditions used in the study 
are thought to be unfavorable for 
 {\it Physarum} plasmodium, the spontaneous slowdown can be interpreted 
as anticipated behaviors against unfavorable conditions
\citep{sai08}.

In this paper, we present mathematical models   
to explain this period-memorizing behavior.
Since there is little information about  
the controlling mechanisms of the {\it Physarum} plasmodium's
locomotion, we assume a general dynamical system,
and examine the basic characteristics required to  
control the plasmodium's locomotion speed.
We then present a simple model that 
fulfills these requirements and qualitatively reproduces 
experimental evidences. 
We insist the model is {\it minimal}
in the sense that it has the minimum degrees of freedom and 
that the functions that it uses 
(including the differential equations to drive the system) 
are all linear functions. 
We also propose some modifications of the minimal model to
improve the reproducibility  of experimental evidence.
Another mathematical model has been presented 
in a paper reporting these experiments \citep{sai08}, 
though that model's mechanism differs from ours.
We outline the previous model and compare it with our model
in the Discussion section.

\section{Basic requirements for models} 

In this section, we argue the class of models for 
a biochemical system 
which controls the reduction of the locomotion speed of the
{\it Physarum} plasmodium.
The main characteristics of the system are that 
the stimulation makes the system reduce the locomotion speed, and 
the system also spontaneously reduces the speed.
Given that the speed reduction is a transient response,
the systems behaviors are analogous to those of a neuron with 
steady and pulsing inputs.
Thus, our model refers to simple neuron models: 
integrate-and-fire models \citep{bur06}.
We consider differential equations 
\begin{eqnarray}
\dot{x}_i=g(x_1,x_2,\cdots,x_n), \ \ i=1,\cdots,n, \label{eq:0} 
\end{eqnarray}
to describe the autonomous dynamics of the biochemical system in 
{\it P.  polycephalum}, 
and suppose that this system reduces 
the locomotion speed when the 
variables satisfy a condition 
\begin{eqnarray}
f(x_1,x_2,\cdots,x_n)>0. \label{eq:1} 
\end{eqnarray}
In other words, $f(x_1,x_2,\cdots,x_n)=0$ determines the threshold 
plane, and if the system gets across the plane, 
locomotion speed is reduced.
If the stimulation is imposed, the orbit of the system in the 
sub-threshold region is forced to shift into the over-threshold region.
On the other hand, if the orbit spontaneously crosses over the 
the threshold plane, SPS occurs.

In the following, 
we examine the experimental evidences,  
list four basic requirements for models, 
and give appropriate assumptions to meet these requirements.

{\bf Equilibrium states.}
Before the periodic stimulation was applied or long after
stimulation was ceased, the plasmodium showed steady locomotion.
Thus, the system should have equilibrium states in the
sub-threshold region ($f(x_1,x_2,\cdots,x_n)<0$),
so that it does not reduce speed spontaneously 
in the  equilibrium states.

{\bf Locomotion recovery.}
{\it Physarum} plasmodium recovered its locomotion speed after 
stimulation,
and became ready to respond to the next stimulation.
This implies that the condition eq. (\ref{eq:1}) does not hold 
for long after the slowdown begins.
In order to bring the system's state back to the sub-threshold region,
here we assume that some of the variables are 
reset after the system induces the slowdown.
Let $x_1,\cdots, x_m$ be the variables to be reset
and $y_1=x_{m+1},\cdots, y_{n-m}=x_n$ the variables not to be reset
 after the slowdown.

{\bf Transitive spontaneous action.} 
SPS did not persist in the experiments.
Thus, 
there should be a difference in the system's actions between 
a stimulation-induced slowdown and spontaneous slowdown (SPS).
We assume that the resetting manners differ.
In the SPS case, the reset may start before 
the orbit invade deeply into the over-threshold region.
However, the stimulation can be vigorous and may shift 
variables widely so that the function (\ref{eq:1}) 
takes a large positive value.
This value may affect the resetting manner.
In particular, we assume that  after the stimulation,
resettable variables are shifted to the complete reset state
\begin{eqnarray}
x_i
\stackrel{\textrm{\scriptsize reset}}{\longrightarrow}
\overline{x}_i, \ \ i=1,\cdots,m. \label{eq:2} 
\end{eqnarray}
On the other hand, the reset after SPS is 
incomplete: resettable variables are shifted to an internally dividing 
state between the state just before the reset and the complete reset 
state
\begin{eqnarray}
x_i
\stackrel{\textrm{\scriptsize reset'}}{\longrightarrow}
\lambda_i x_i + (1-\lambda_i) \overline{x}_i, 
\ \ i=1,\cdots,m.  \label{eq:3}
\end{eqnarray}

{\bf Memory.}
The plasmodium remembered the stimulation period for a long time 
after stimulation stopped.
In order for a dynamical system to 
memorize such an analog value, 
it must have a slow manifold in the phase space:
the relaxation along with the manifold is quite slow so that 
the system can preserve the information about its own history 
for a long time \citep{fox02}.
As a simple and extreme case, we introduce the neutral manifold 
(a line of equilibrium states) 
in the phase space of our model, where the system never relaxes in that  
direction, and thus, the stored memory is permanently preserved.
In other words, we introduce a conserved quantity whose value
 never changes in the autonomous dynamics.
Along with those dynamics, the system's orbit is 
constrained on a $n-1$ dimensional invariant plane 
transverse to the neutral manifold, and 
the resetting action will shift the orbit to another plane 
along with the neutral manifold.
The period of stimulations is thought to control which 
plane is selected.

Summarizing the class of models,  we can say that these models 
have a conservative quantity; thus the autonomous dynamics of the 
models are restricted to the $n-1$ dimensional invariant plane.
In this plane there is an equilibrium state, a $n-2$ dimensional 
threshold plane, and a complete reset state. 
Changing the value of the conservative quantity
changes the positional relation among them in the plane.
The relation determines whether or not autonomous dynamics from the 
complete reset state to the equilibrium state cross 
the threshold plane, and if they do, it also determines the 
period of time from the complete reset
state to the threshold plane.

\section{Models and their behaviors} 

Based on the above discussions,  
here we present a minimal model for 
the slowdown induction system. 
The model has two resettable variables $x_1$, $x_2$, 
and one un-resettable variables $y$, 
driven by linear differential equations, 
\begin{eqnarray}
\left\{
\begin{array}{l}
~~~\dot{x}_1=x_2-x_1\\
\tau_x\dot{x}_2=y-x_2\\
~\tau_y\dot{y}=x_1-y
\end{array}
\right. \label{eq:4}
\end{eqnarray}
where $\tau_x$ and $\tau_y$ indicate the relative time scales of 
$x_2$, and $y$ respectively.
These equations have a conserved quantity
\begin{eqnarray}
x_1+\tau_x x_2+\tau_y y ={\rm constant}. \label{eq:5}
\end{eqnarray}
Thus there are neutral manifold in the phase space, 
and the equilibrium states form a line 
\begin{eqnarray}
x_1=x_2=y.  \label{eq:6}
\end{eqnarray}
We set the threshold function as  
\begin{eqnarray}
f(x_1,x_2,y)=x_1-y-\delta, \label{eq:7}
\end{eqnarray}
where $\delta$ determines the distance between the 
equilibrium state and the threshold plane.
We choose the complete reset state for resettable variables
\begin{eqnarray}
(\overline{x}_1,\overline{x}_2)=(0,1).  \label{eq:8}
\end{eqnarray}
Thus, reset after stimulation becomes 
\begin{eqnarray}
\left\{
\begin{array}{l}
x_1 \stackrel{\textrm{\scriptsize reset}}{\longrightarrow}
 0\\
x_2 \stackrel{\textrm{\scriptsize reset}}{\longrightarrow}
 1
\end{array}
\right. \label{eq:9}
\end{eqnarray}
and reset after SPS becomes  
\begin{eqnarray}
\left\{
\begin{array}{l}
x_1\stackrel{\textrm{\scriptsize reset'}}{\longrightarrow}
 \lambda_1\cdot x_1\\
x_2\stackrel{\textrm{\scriptsize reset'}}{\longrightarrow}
 \lambda_2\cdot x_2+(1-\lambda_2)
\end{array}
\right. \label{eq:10}
\end{eqnarray}

\begin{figure}
\includegraphics[width=11.7cm]{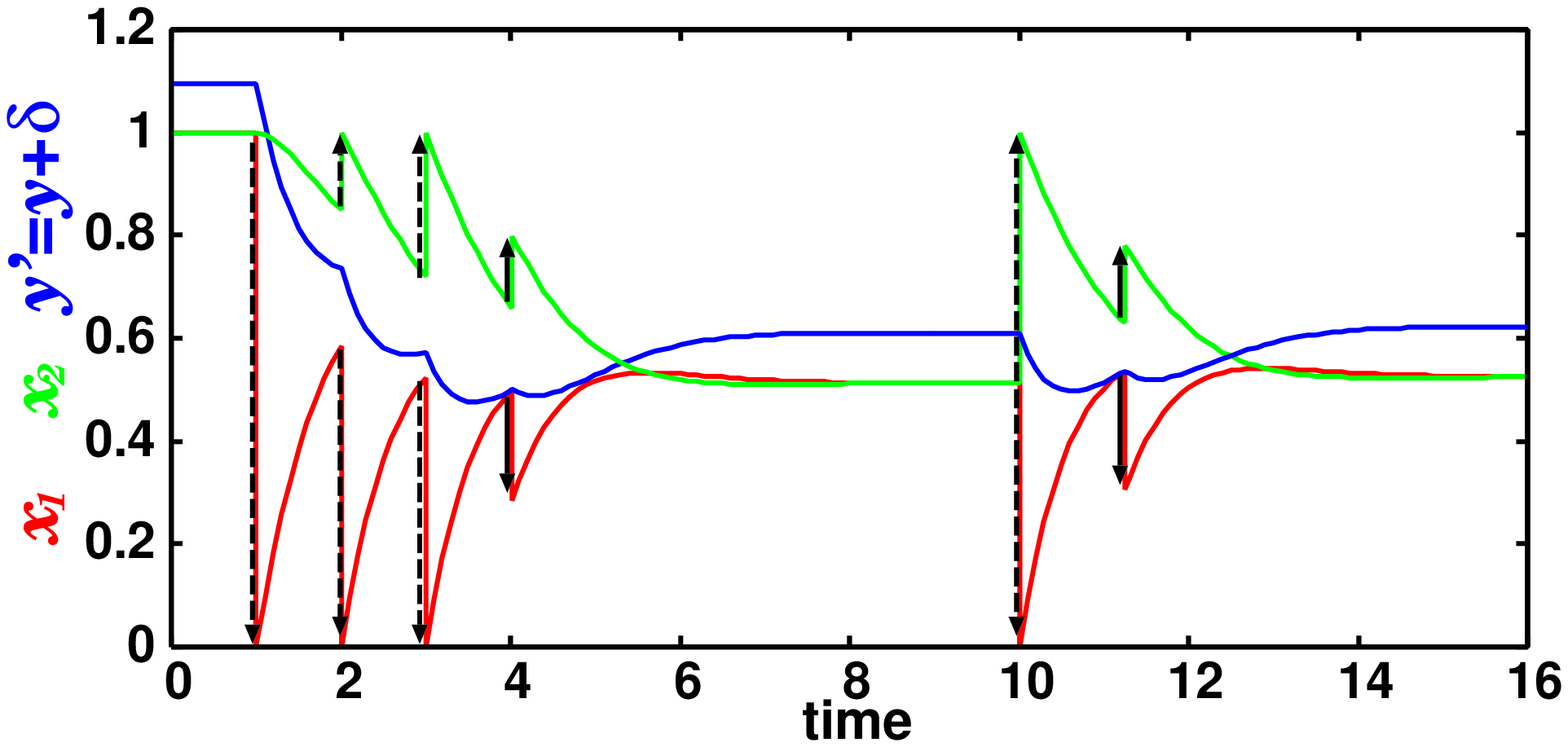}
\caption{
Time series of the minimal linear model. 
$(\tau_x,\tau_y,\delta,\lambda_1,\lambda_2)=
(1.11,1.23,0.0961,0.571,0.592)$ are used. 
Red line indicates the time series of $x_1$, 
green line indicates $x_2$, and blue line 
indicates $y'=y+\delta$ instead of $y$ 
(so that spontaneous crossing over of the threshold 
is identified by crossing of red and blue lines).
Stimulations are induced at $t=1,2,3,10$.
Arrows with dashed lines indicate the reset of 
resettable variables after the stimulations, 
and arrows with solid lines indicate the reset by 
the spontaneous crossing over the threshold.}
\label{fig:1}
\end{figure}

We show the behavior of our model in fig. \ref{fig:1}.
Initially the system stays at the equilibrium point $x_1=x_2=y=1$.
We give the stimulation three times. 
After that, 
at $t=4.03$, 
the system reaches the point where eq. (\ref{eq:7}) changes the sign 
from negative to positive, SPS occurs,
and the orbit relaxes to a new equilibrium  point  $x_1=x_2=y\simeq0.51$.
After additional stimulation is imposed ($t=10$), 
SPSD occurs at $t=11.26$.
and the orbit relaxes to an equilibrium point that is closer to  the 
previous one ($x_1=x_2=y\simeq0.52$).

\begin{figure}
\includegraphics[width=6cm]{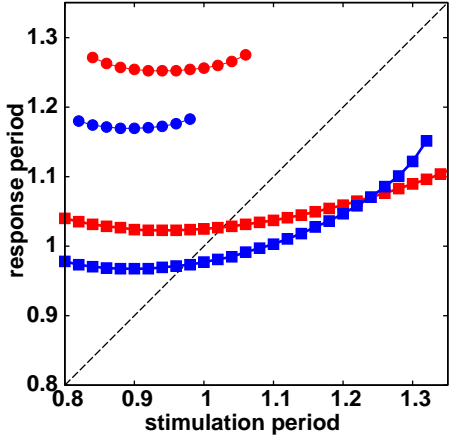}
\caption{
Response periods after various periods of stimulations.
Data from minimal linear model 
and from nonlinear model  (in which  
eq. (\ref{eq:11}) and 
(\ref{eq:12}) are used instead of  eq. (\ref{eq:4}) and 
(\ref{eq:10}) ) are plotted with red and blue respectively,
with parameters 
$(\tau_x,\tau_y,\delta,\lambda_1,\lambda_2)=
(1.11,1.23,0.0961,0.571,0.592)$ for the linear model and
$(\tau'_x,\tau'_y,\delta',\lambda_1,\lambda_2,
\omega_{x1},\omega_{x2},\omega_{x3}
\omega_{y1},\omega_{y2},\omega_{y3}
\omega_{\delta1},\omega_{\delta2},\omega_{\delta3})=$
$(0.926,1.32,0.0672,0.46,0.805,0.116,-0.111,-0.0803,-0.118,0.151,-0.089,
0.047,0.0603,0.0577)$ for the nonlinear model.
The lengths of time between the last stimulation of periodic
stimulation and SPS are plotted with filled squares, and
the lengths of time between the one time stimulation and 
and SPSD are plotted with filled circles.
}
\label{fig:2}
\end{figure}

We repeated the simulations with various stimulation periods.   
The response periods of SPS and SPSD 
are shown in fig. \ref{fig:2}.
The parameters used in these simulations are tuned so that 
these response periods have similar values to the stimulation periods 
and show positive correlations as the stimulation period changes.

\begin{figure}
\includegraphics[width=13cm]{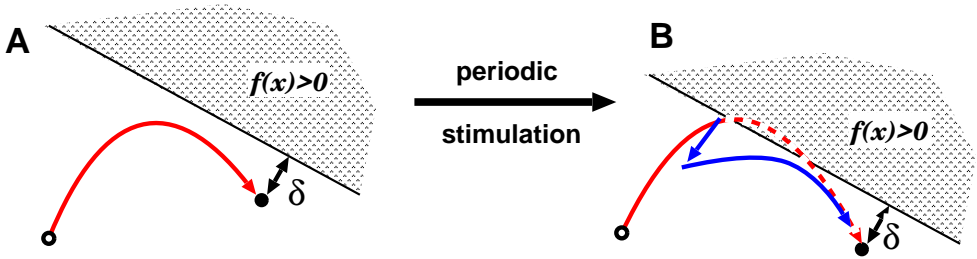}
\caption{
The schematic figures of the invariant planes transverse to the 
neutral manifold and orbits on them.
Plane A (displayed at left) corresponds to initial unstimulated 
conditions, and plane B (right) corresponds to 
the conditions after the periodic stimulation ceases.
The white area indicates the sub-threshold region, and 
the shaded area indicates the over-threshold region.
The black open circle indicates the complete reset state, and 
the black filled circle indicates the equilibrium state.
The red arrow shows the relaxation orbit from 
the complete reset state to the equilibrium state.
In the plane B, the orbit crosses the threshold line
(the subsequent trajectory is drawn by dashed line).
The orbit then shifts a bit toward the complete reset 
state, and relaxes to the equilibrium state. 
Both are drawn by blue arrows.
}
\label{fig:3}
\end{figure}

Here we sketch the spontaneous dynamics of the system, 
and verify that the above model with three 
variables is a minimal model.
Because the system has one conserved quantity, 
the spontaneous dynamics are restricted in a two dimensional 
plane, which is described in fig. \ref{fig:3}.
The plane is divided by threshold line eq. (\ref{eq:7}) 
and has one equilibrium  point and one complete reset state
which lies in the sub-threshold region.
Initially the arrangement of the plane may look like
fig. \ref{fig:3}-A.
The relaxation orbit from the complete reset state 
directly relaxes to the equilibrium  point. 
Periodic stimulation  may change 
the positional relationship among 
the equilibrium  point, the complete reset state, 
 and the sub-threshold line, similar to fig. \ref{fig:3}-B.
That occurs due to the continuous change of the unresettable variable 
during the periodic stimulation.
In the plane, the orbit from the complete reset state
violates the threshold line in the course of relaxation 
to the equilibrium point, 
and thus SPS occurs.
After the induction, the orbit shifts a bit toward the 
the complete reset state and relaxes to the equilibrium  point.
We emphasize that after periodic stimulation, 
the orbit from the complete reset state 
overreaches and crosses the threshold line, but 
the orbit after spontaneous induction directly 
relaxes to the equilibrium  point.
This happens because the plane is two-dimensional.
If it is one-dimensional, there are only two cases: 
if the equilibrium point is in the sub-threshold region, the orbit 
relaxing to it never crosses the threshold, and 
if the equilibrium  point is in the over-threshold region  
(this case is eliminated in the above assumptions),
the orbit never reaches the equilibrium  point.
This ensures that the plane should be at least two-dimensional.
Thus, with the neutral manifold for memory, a minimal model
must have three variables.

To make the model minimal and easily understandable, 
we chose linear functions for differential equation eq. (\ref{eq:4}),
the threshold function eq. (\ref{eq:7}) and resetting manner
eq. (\ref{eq:10}).
However, these functions can be nonlinear without changing the
basic mechanisms.
Here we give an example of the nonlinear modification of the model, 
by letting parameters $\tau_x$, $\tau_y$ 
depend on variables
\begin{eqnarray}
\left\{
\begin{array}{l}
\displaystyle
\frac{1}{\tau_x}\to\frac{1}{\tau'_x}\cdot \left(1+\omega_{x1}\cdot x_1+
\omega_{x2}\cdot x_2+\omega_{x3}\cdot y\right)\\
\displaystyle
\frac{1}{\tau_y}\to\frac{1}{\tau'_y}\cdot \left(1+\omega_{y1}\cdot x_1+
\omega_{y2}\cdot x_2+\omega_{y3}\cdot y\right)
\end{array}
\right. \label{eq:11}
\end{eqnarray}
 and changing the threshold function from eq. (\ref{eq:7}) to
\begin{eqnarray}
f_{\rm nlin}(x_1,x_2,y)=x_1-y\cdot \left(1+\omega_{\delta1}\cdot x_1
+\omega_{\delta2}\cdot x_2+\omega_{\delta3}\cdot y\right)-\delta
\label{eq:12}
\end{eqnarray}
where $w_i$ are the nonlinearity parameters. 
The modification improves the reproducibility of 
the periods, as displayed in fig. \ref{fig:2}.

\begin{figure}
\includegraphics[width=6.7cm]{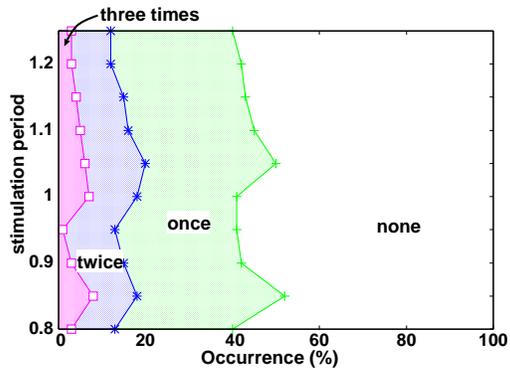}
\caption{
Number of SPS responses in the linear model
with Langevin noise, eq. (\ref{eq:12}).
$(\tau_x,\tau_y,\delta,\lambda_1,\lambda_2)=
(1.11,1.23,0.1361,0.571,0.592)$ and  
$\sigma=0.04$ are used for the simulation.
For each stimulation  period, 
100 simulation runs are performed and 
statistics on the number of SPS responses are taken.
The magenta area represents the 
proportion of cases showing three SPS responses,
blue shows  the proportion of cases with two SPS responses,
and green and white show the single-response and no-response 
proportions, respectively.
}
\label{fig:4}
\end{figure}

In experiments, the number of times of SPS occurrences varied among;
whereas one organism showed SPS three times, 
another organism did not at all. 
These variations must be due to fluctuation in the cell state 
dynamics.
Adding Langevin noise to the differential equations 
\begin{eqnarray}
\left\{
\begin{array}{l}
~~~\dot{x}_1=x_2-x_1 +\sigma\xi_1(t) \\
\tau_x\dot{x}_2=y-x_2 +\sigma\xi_3(t) \\
~\tau_y\dot{y}=x_1-y +\sigma\xi_3(t) \\
\end{array}
\right. \label{eq:13}
\end{eqnarray}
where $\langle
\xi_i(t_1)\xi_j(t_2)\rangle=\delta_{ij}\cdot\delta(t_1-t_2)$,
our model also shows such varieties of responses, 
as shown in fig. \ref{fig:4}.

\section{discussion} 

In this paper, we presented a mathematical model 
for the period-memorizing behavior of the plasmodium 
of {\it P. polycephalum}.
We described and examined the conditions for 
satisfying the model,  
and introduced a minimal linear model that 
qualitatively reproduces 
the experimental evidence.
To improve the model, we also suggests the two methods 
(nonlinearization and noise addition) that  
are better for quantitative reproducibility.

Since almost nothing has been revealed about the 
chemical components for the locomotion speed reduction system 
of the {\it Physarum} plasmodium, 
we did not present a detailed model in which variables 
 correspond to 
actual chemical components.
Instead, we made a minimal linear model whose mechanism is 
easily comprehensible. 
In our model, for example, the system's resetting action after it reduces 
the locomotion speed is modeled in a conditional branching manner, 
instead of in an autonomous dynamical system.
In a detailed model, this process may be represented by neuron-like 
firing action. 
The abstraction revel of our model corresponds to 
those of integrate-and-fire models for a neuron \citep{bur06},
as mentioned above.
The transcriptome of {\it P. polycephalum} is now being sequenced
\citep{glo08}.
This information may lead us to develop a more detailed model. 
In particular, two important actions for locomotion, actomyosin 
contraction and sol-gel transition of the cytoplasm, are thought to 
be controlled by the calcium ion concentration.
Thus, transcripts homologous to what are involved in 
 calcium ion concentration regulations in other organisms
may give useful insight into the detailed mechanism.

Previously, another model for the plasmodium's period-memorizing 
behavior was presented by \cite{sai08}.
Here we outline that model's architecture and behaviors.
That previous model consists of decoupled multiple oscillators. 
The oscillators form sub-groups, 
each of which  contains a certain number of oscillators
with the same frequency ($\propto 1/{\rm period}$). 
There are multiple sub-groups with different frequencies.
The phase relations among oscillators are random at the initial condition.
The periodic stimulation first synchronizes oscillators in phase 
within sub-groups that have similar periods to those in the stimulation,
and then it synchronizes these sub-groups in phase.
In that study, this super cluster (a cluster of clustered sub-groups)
is assumed to induce SPS in their study.
After stimulation is ceased,  the super cluster 
gradually desynchronizes and SPSs are observed 
as only transient phenomena.
However, the clusters within the sub-groups do not desynchronize, 
because they have exactly the same periods.
An additional one-time stimulation again induces the formation of 
a super cluster, and thus induce SPSD. 
In the following, we remark on the two advantages 
of our model in reproduction of experimental evidence, 
and discuss two other prominent differences in the mechanism
 between the previous model and our model.

With the previous model, the occurrence of SPSD depends 
on at which phase the one-time stimulation is given; 
stimulation at the in-phase with previous periodic stimulation 
induces SPSD much more strongly than
the stimulation at the anti-phase. 
However, this is not consistent with 
the experimental results. 
In the experiments, the occurrence of SPSD did not depend
on a small shift in the timing of one-time stimulation
\footnote{the occurrence of SPSD  seems to decrease 
monotonically when the delay in timing increases.
This  may be explained by the noise effect in both models.
}, which is reproduced in our model. 
Besides, SPSD was observed only at $T=60$ min in the experiments, 
and our model also gives SPSD in a certain window of the 
stimulation period
($T=0.84\sim1.06$ with the linear model and $T=0.82\sim0.98$ with 
the nonlinear model).  
However, the occurrence of SPSD in the previous model does not 
depend on the stimulation period. 
Thus another mechanism will be required for 
the previous model to reproduce 
the occurrence of SPSD for a certain window of the stimulation  period.

There are large differences in required degrees of freedom,
between our model and the previous model.
As described above, a response period in the previous model 
is represented by intrinsic collective modes of the dynamical system.
This means that a number of variables (oscillators) are 
needed to respond to each stimulation period.
Thus the model needs a lot of degrees of freedom.
Actually, $4.4\times 10^{5}$ oscillators are used in the 
numerical simulation in the paper \citep{sai08}:
it contains 440 sub-groups with different frequencies 
and each sub-group consists of 1000 oscillators. 
In our model, on the other hand, a response period is not generated 
by the intrinsic mode but is represented by the length in phase space 
between the reset state and the threshold plane, which is 
continuously adjustable.
Thus our model shows qualitatively the same behaviors with only  
three degrees of freedom.

Because the previous model  uses intrinsic modes, 
the stimulation period and response period 
show fairly good correspondence.
No tuning of parameters is necessary for the correspondence. 
On the other hand, in our model, parameter tuning is necessary
even for reproducing the qualitative behaviors.
It should be noted that having a conserved quantity in a general 
dynamical system itself demands the tuning of parameters, 
since the system is structurally unstable \citep{guc83}.  
However, if this period-memorizing qualification has an adaptive role 
as means for anticipating unfavorable conditions, 
it might be not so hard for biological organisms to tune these 
parameters in evolution.
For example, the period and the response to the temperature changes of 
circadian oscillation systems are thought to be tuned evolutionarily 
in many organisms \citep{ros97}.

\section{Acknowledgments}

The author is grateful to T. Nakagaki, A. Awazu, and S. Ishihara for 
providing the author with motivation to conduct this study.

\end{document}